\documentclass[aps,twocolumn]{revtex4}

\begin{document}

\newcommand{\dfrac}[2]{\frac{\displaystyle #1}{\displaystyle #2}}
\renewcommand{\thepage}{VPI-IPPAP-02-09}


\title{The NuTeV Anomaly, Neutrino Mixing, and a Heavy Higgs}
\author{Tatsu Takeuchi}
\affiliation{IPPAP, Physics Department, Virginia Tech, Blacksburg, VA 24061,
U.S.A.\\
Talk presented at the YITP workshop `Progress in 
Particle Physics', July 9, 2002.
}

\maketitle


\thispagestyle{headings}


The NuTeV experiment \cite{Zeller:2001hh} 
has measured the ratios of neutral to charged current events in 
muon (anti)neutrino -- nucleon scattering:
\begin{eqnarray}
R_\nu 
& = & \dfrac{ \sigma(\nu_\mu N \rightarrow \nu_\mu X) }
            { \sigma(\nu_\mu N \rightarrow \mu^-   X) }
\;=\; g_L^2 + r g_R^2\;, \cr
R_{\bar{\nu}}
& = & \dfrac{ \sigma(\bar{\nu}_\mu N \rightarrow \bar{\nu}_\mu X) }
            { \sigma(\bar{\nu}_\mu N \rightarrow \mu^+         X) }
\;=\; g_L^2 + \dfrac{g_R^2}{r}\;,
\end{eqnarray}
where
\begin{equation}
r = \dfrac{ \sigma( \bar{\nu}_\mu N \rightarrow \mu^+ X) }
          { \sigma( \nu_\mu       N \rightarrow \mu^- X) }
\sim \frac{1}{2}\;,
\end{equation}
and has determined the parameters $g_L^2$ and $g_R^2$ to be
\begin{eqnarray}
g_L^2 & = & 0.3005 \pm 0.0014\;, \cr
g_R^2 & = & 0.0310 \pm 0.0011\;.
\label{nutev}
\end{eqnarray}
The Standard Model (SM) predictions based on a fit to
LEP/SLD data are cited as
$[g_L^2]_\mathrm{SM}=0.3042$ and $[g_R^2]_\mathrm{SM}=0.0301$
in Ref.~\cite{Zeller:2001hh}
and we see a $2.6\sigma$ disagreement in $g_L^2$.
In terms of the ratios $R_\nu$ and $R_{\bar{\nu}}$, this means that the 
neutral current events are not as numerous as the SM predicts
when compared to the charged current events.
On the LEP/SLD side, the measured invisible width is also known 
to be $2\sigma$ below the SM prediction.
Both these facts seem to suggest that the
$Z\nu\nu$ coupling is somehow suppressed.

Suppression of the $Z\nu\nu$ coupling can be arranged in 
a variety of models by mixing the neutrino with a heavy
gauge singlet state as discussed in Ref.~\cite{Davidson:2001ji}.
For the sake of simplicity, consider the case 
where all three generations of light neutrinos mix with a universal angle 
with such states.  In such theories, if the
$Z\nu\nu$ couplings are suppressed by a universal factor of 
$(1-\varepsilon)$, then the $W\ell\nu$ couplings are also suppressed 
by a factor of $(1-\varepsilon/2)$.  Then the numerators of $R_\nu$ and
$R_{\bar{\nu}}$ will be suppressed over their denominators, so
such a mixing could in principle explain the NuTeV anomaly.

However, one must recall that one of the inputs used in calculating the
SM predictions is the Fermi constant $G_F$
which is extracted from the muon decay constant $G_\mu$.
The suppression of the $W\ell\nu$ couplings would lead to the correction
\begin{equation}
G_F = G_\mu (1+\varepsilon)\;,
\end{equation}
which will affect all SM predictions.
The authors of Ref.~\cite{Davidson:2001ji}
conclude that $\varepsilon$ cannot be chosen to explain the 
NuTeV data without destroying the excellent agreement between
LEP/SLD and the SM.

But what if we include oblique corrections from new physics 
\cite{Peskin:1990zt} to compensate for the shift in $G_F$?  
Note that the Fermi constant always appears multiplied by 
the $\rho$-parameter in neutral current amplitudes 
so a shift in $G_F$ should be absorbable into the $T$-parameter.

In Ref.~\cite{LOTW}, 
we perform a global fit to the NuTeV and LEP/SLD data
to determine what values of $\varepsilon$, $S$, and $T$ are
preferred.
Using $m_H = 115\,\mathrm{GeV}$, $m_t = 174.3\,\mathrm{GeV}$
as the reference SM, we obtain excellent agreement between
theory and experiment at
\begin{eqnarray}
S & = & -0.05\pm 0.10\;, \cr
T & = & -0.40\pm 0.16\;, \cr
\varepsilon & = & 0.0028 \pm 0.0010\;.
\end{eqnarray}
What kind of `new physics' is compatible with these
values of $S$ and $T$?  It turns out that the simplest
solution is the SM itself with a larger Higgs mass.
Thus neutrino mixing together with a heavy Higgs will reconcile
the LEP/SLD and NuTeV results.

Details of this analysis will be presented in Ref.~\cite{LOTW}.

\smallskip
\noindent
\textbf{\small Acknowledgments}

I would like to thank 
Will~Loinaz, Naotoshi~Okamura, and L.~C.~R.~Wijewardhana for
their collaboration on this project, and 
Michio~Hashimoto and Mihoko~Nojiri
for helpful discussions.
This research was supported by the U.S. Department of Energy, 
grant DE--FG05--92ER40709, Task A.


%
\end{document}